\documentclass{easychair}

\usepackage{microtype}
\usepackage[T1]{fontenc}
\usepackage{lmodern}
\usepackage[utf8]{inputenc}
\usepackage{siunitx}
\usepackage{booktabs}
\usepackage{tabularx}
\usepackage{amssymb}
\usepackage{tikz}
\usepackage{listings}
\usepackage[toc,page]{appendix}
\usepackage{subcaption}
\usepackage[font=footnotesize,labelfont=bf]{caption}
\usepackage[noabbrev,nameinlink,capitalize,poorman]{cleveref}

\title{Reset Early, Reset Often, Eliminate Models}
\author{Martin S.\ Fixman\inst{1} \and Fredrik Rømming\inst{2} \and Sean B.\ Holden\inst{3}}

\institute{
	University of Cambridge, \\
	United Kingdom \\
	{}\inst{1}\email{msf43@cl.cam.ac.uk} \qquad\quad {}\inst{2}\email{fr409@cl.cam.ac.uk} \qquad\quad {}\inst{3}\email{sbh11@cl.cam.ac.uk}
}

\titlerunning{Reset Early, Reset Often, Eliminate Models}
\authorrunning{Fixman, Rømming, Holden}

\graphicspath{{figures/}}

\hypersetup{
      pdftitle={Reset Early, Reset Often, Eliminate Models},
      pdfauthor={Martin S. Fixman, Fredrik Rømming, Sean B. Holden}
}

\lstdefinelanguage{pseudocode}{
	keywords={python,is,true,unknown,false,any,all,in,and,or,def,while,if,else,return,continue,restart,step},
	sensitive=true,
}

\lstdefinestyle{pseudocode}{
	aboveskip=0.75\baselineskip,
	backgroundcolor=\color{black!4},
	basicstyle=\ttfamily\scriptsize,
	belowskip=0.75\baselineskip,
	breakatwhitespace=true,
	breaklines=true,
	captionpos=b,
	columns=fullflexible,
	commentstyle=\itshape\color{black!55},
	framerule=1pt,
	framesep=5pt,
	frame=single,
	framexbottommargin=0.25em,
	framexleftmargin=0.4em,
	framexrightmargin=0.4em,
	framextopmargin=0.25em,
	keepspaces=true,
	keywordstyle=\bfseries\color{blue!70!black},
	language=pseudocode,
	numbers=left,
	numberstyle=\tiny,
	rulecolor=\color{black!22},
	showstringspaces=false,
	tabsize=2,
	xleftmargin=1.25em,
	xrightmargin=1.25em,
}
\newcommand{\fredrik}[1]{}
\newcommand{\fredriksolved}[1]{}

\newcommand{\step}[1]{\textsf{#1}}
\newcommand{\cop}[1]{\textsf{#1}}

\newcommand{\leancopc}{\cop{leanCoP\textsuperscript{Con}}}
\newcommand{\satcopzc}{\cop{SATCoP\textsubscript{0}\textsuperscript{Con}}}
\newcommand{\satcopc}{\cop{SATCoP\textsuperscript{Con}}}

\crefname{appendix}{Appendix}{Appendices}
\Crefname{appendix}{Appendix}{Appendices}

\begin{document}
\maketitle{}

\begin{abstract}
	Connection-tableau provers typically construct closed tableaux by backtracking search.
	This gives lightweight proof procedures, but their low-memory search state leaves little derived information to reuse after a failed attempt.
	\cop{SATCoP} addresses this by retaining grounded tableau clauses in an incremental SAT solver, allowing the prover to refute the problem once the accumulated ground clauses become propositionally unsatisfiable.
	We introduce \cop{SATResetCoP}, which relies more directly on this persistent SAT state by not backtracking on unproductive tableaux and instead starting new ones to generate further ground clauses for the SAT solver.
	On TPTP, \cop{SATResetCoP} improves over our \cop{leanCoP}-style baseline by 54\% and our \cop{SATCoP}-style baseline by 18\%.
	On the bushy \cop{MPTP2078} benchmark, it solves \num{616} problems, a 49\% improvement over our \cop{leanCoP}-style baseline and a 12\% improvement over our \cop{SATCoP}-style baseline.
	These results suggest that changing tableau guidance to accumulate useful ground clauses faster improves the performance of the prover.
\end{abstract}


\section{Introduction}
Connection-tableau provers prove first-order problems by constructing closed tableaux from sound tableau-construction rules~\cite{connectiontableaux}.
Implementations usually turn this construction into backtracking search over applications of those rules.
This gives a lightweight and goal-directed proof procedure, but its small search state means that failed partial tableaux leave little derived information for later attempts to reuse.

Much practical work on connection-tableau proving has therefore focused on controlling backtracking, and learning how to restrict it.
The line from semantic tableaux and model elimination to \cop{leanTAP} and \cop{leanCoP} showed that goal-directed tableau search could be implemented in small, effective proof procedures~\cite{semantictableaux,modelelimination,leantap,leancop}.

\cop{leanCoP} uses Prolog's SLD resolution for depth-first connection-tableau search with iterative deepening, and \cop{leanCoP 2.0} obtains substantial speedups by restricting this backtracking with Prolog cuts at the cost of completeness~\cite{leancopcut}.
\cop{meanCoP} refines this pruning strategy by separating cuts on reductions from cuts on extensions and by introducing exclusive cuts, which preserve more alternatives than the inclusive cuts used by \cop{leanCoP}~\cite{meancop}.

\cop{SATCoP} addresses the same weakness from a different direction~\cite{satcop}.
During tableau construction, it grounds encountered clauses and stores them in an incremental SAT solver.
If the accumulated ground clauses become propositionally unsatisfiable, the original first-order problem can be proved before the active tableau has closed.
We introduce \cop{SATResetCoP}, which pushes this idea further: when the active tableau stops producing new ground clauses, it is reset, while the accumulated SAT solver state is kept.
The tableau is therefore used as a generator of ground clauses, and the SAT solver becomes the persistent object from which a proof may ultimately be obtained.

We evaluate \cop{SATResetCoP} on TPTP against \cop{leanCoP}-style and \cop{SATCoP}-style baselines implemented in a fully compatible backend.
\cop{SATResetCoP} solves 54\% more problems than the \cop{leanCoP}-style baseline and 18\% more than the \cop{SATCoP}-style baseline, suggesting that once ground clauses are preserved in the SAT solver, starting a new tableau rather than continuing to backtrack on an unproductive one can be an effective way to generate the clauses needed for a proof.

\section{Preliminaries}

\subsection{Connection Calculus and Tableau Search}

Connection tableaux and model elimination give a goal-directed alternative to the saturation style used by provers such as \cop{Vampire}~\cite{riazanovDesignImplementationVAMPIRE2002}.
Instead of deriving clauses from a global clause set until a contradiction is found~\cite{bachmairResolutionTheoremProving2001,schulzPerformanceClauseSelection2016}, connection calculi try to close a tableau by connecting goal literals to complementary literals, either on the current path or in a fresh copy of an input clause~\cite{connectiontableaux,ottenAdvancesConnectionBasedAutomated2017}.

The standard clausal connection tableau has three proof-building rules~\cite{connectiontableaux,ottenSyntaxConnectionProofs2023}.
\begin{description}
	\item[\cop{Start}] Choose an input clause and make its literals open goals.
	\item[\cop{Extension}] Open a branch on a clause containing a complementary literal, and add the remaining literals as new subgoals.
		This may require a substitution of variables with terms.
	\item[\cop{Reduction}] Close the current goal against a complementary literal already on the current branch.
		This may also require a substitution.
\end{description}

The rules are sound, and with the possibility of backtracking over rule applications the calculus is complete.
The calculus is not confluent: a locally sound connection can bind variables in a way that prevents later branches from closing~\cite{connectiontableaux}.
Backtracking is therefore part of the operational search procedure, not a proof-building rule.


\begin{figure}[h]
	\centering
	\framebox[\linewidth][c]{\newcommand{\satdistance}{-160pt}
\newcommand{\dist}{65pt}
\newcommand{\lat}{10pt}
\newcommand{\bet}{5pt}

\begin{tikzpicture}[
  attempt/.style = {
    every node/.style = {
      draw,
      rounded corners=1pt,
      inner xsep=3pt,
      inner ysep=2pt,
      font=\footnotesize
    },
    level distance=40pt,
    sibling distance=40pt,
    level 2/.style={sibling distance=63pt},
    edge from parent/.style={draw, -}
  },
  start/.style={},
  goal/.style={fill=black!40},
  bt/.style={edge from parent/.style={draw, dash pattern=on 2pt off 1pt on .5pt off 1pt}},
  title/.style={draw=none, font=\normalsize\bfseries},
  edgelabel/.style={draw=none, fill=none, font=\tiny}
]
\useasboundingbox (-85pt,-122pt) rectangle (407pt,11pt);

\begin{scope}[attempt]
  \node[title] {\step{Start}};
	\node[start] at (0pt, -28pt) {$F(Y) \vee p$};
\end{scope}

\begin{scope}[xshift={1*\dist}, attempt]
  \node[title] {\step{Extension}};
	\node[start] at (0pt, -28pt) {$F(Y) \vee p$}
	child[xshift={-\lat}] {
		node {$ \neg F(X) \vee p $}
		edge from parent node[edgelabel, right] {$ \sigma_0 = \{ X \leftarrow Y \} $}
	};
\end{scope}

\begin{scope}[xshift={2*\dist}, attempt]
  \node[title] {\step{Extension}};
	\node[start] at (0pt, -28pt) {$F(Y) \vee p$}
		child[xshift={-\lat}] { node {$ \neg F(X) \vee p $}
		child {
			node {$\neg p \vee \neg F(a)$}
			edge from parent node[edgelabel, right] {$ \sigma_1 = \sigma_0 $}
		}
		edge from parent node[edgelabel, right] {$ \sigma_0 = \{ X \leftarrow Y \} $}
	};
\end{scope}

\begin{scope}[xshift={3*\dist}, attempt]
  \node[title] {\step{Reduction}};
	\node[start,label={[edgelabel,font=\tiny,xshift=-7pt]300:$ \sigma_2 = \left\{ \begin{gathered} X \leftarrow a \\ Y \leftarrow a \end{gathered} \right\} $}] at (0pt, -28pt) {$F(a) \vee p$}
		child[xshift={-\lat}] { node[goal] {$ \neg F(a) \vee p $}
		child { node[goal] {$\neg p \vee \neg F(a)$} }
	};
\end{scope}

\begin{scope}[xshift={4.25*\dist}, attempt]
  \node[title] {\step{Extension}};
	\node[start,label={[edgelabel,font=\tiny,xshift=10pt]300:$ \sigma_2 = \left\{ \begin{gathered} X \leftarrow a \\ Y \leftarrow a \end{gathered} \right\} $}] at (0pt, -28pt) {$F(a) \vee p$}
		child[xshift={-\lat}] {
			node[goal] {$ \neg F(a) \vee p $}
			child { node[goal] {$\neg p \vee \neg F(a)$} }
		}
		child[xshift={\lat}] {
			node {$ \neg F(a) \vee \neg p $}
		};
\end{scope}

\end{tikzpicture}}
	\caption{Drawing convention for the tableau diagrams, illustrated on steps from \cop{SYN052-1.p}~\cite{Pelletier86}. Each box is a clause instance in the tableau. An edge records the connection that introduced a child clause, with the edge label showing the resulting substitution. Grey boxes mark clauses whose selected open literal has been closed. A \step{Reduction} closes against a complementary literal already on the current path, so it adds no new child.}
\end{figure}
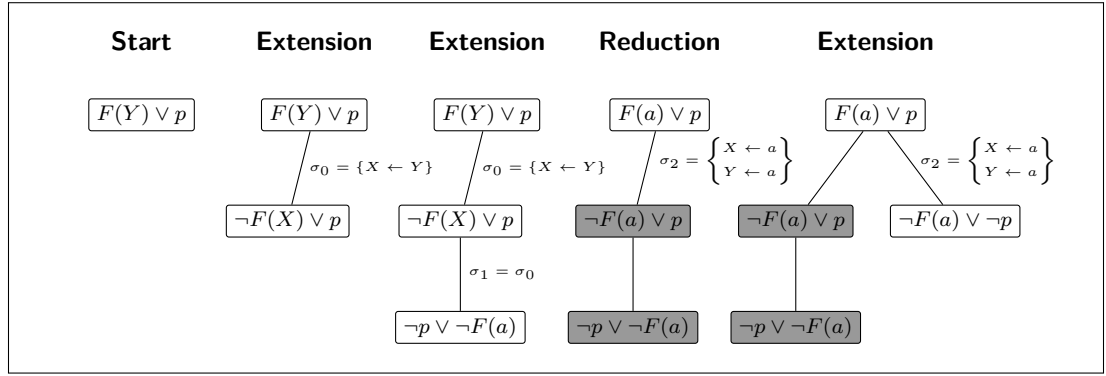

\subsection{Backtracking Control}

Backtracking lets a prover recover from locally sound choices that later prevent tableau closure.
In \cop{leanCoP}-style provers, this search is controlled mainly by iterative deepening and Prolog cuts.
Iterative deepening bounds the tableau height and raises the bound when no proof is found.
Cuts prune alternatives after selected reductions or extensions, reducing search but making the cut phase incomplete unless the surrounding schedule later explores less restricted search~\cite{leancopcut}.

\cop{meanCoP} refines this pruning mechanism by separating cuts on reductions from cuts on extensions and by introducing exclusive cuts, which preserve more alternatives than the inclusive cuts used by \cop{leanCoP}~\cite{meancop}.
\cop{hopCoP} takes a different route: instead of pruning alternatives directly, it learns constraints explaining why a search state is stuck, reducing repeated backtracking while preserving completeness at a fixed iterative-deepening bound~\cite{rawsonConstraintLearningNonconfluent2026}.

\subsection{\cop{SATCoP} and Persistent Ground Clauses}
\label{satcop}

\cop{SATCoP} maintains an incremental SAT solver alongside tableau search~\cite{satcop}.
During tableau search, clauses appearing in the tableau are heuristically grounded and added to the SAT solver.
If the accumulated ground clauses are propositionally unsatisfiable, proof search can stop even if the active first-order tableau has not closed.

The soundness condition is that every clause sent to the SAT solver must be a ground instance of a clause generated by the tableau.
Remaining tableau variables are therefore mapped to ground terms before a clause is added.
Rawson and Reger use the grounding heuristic inspired by a similar suggestion by Schulz~\cite{schulzSATInterpretation}: choose the most frequent constant in the conjecture, falling back to a fresh constant when no conjecture constant is available~\cite{satcop}.

Later \cop{SATCoP} variants also use the SAT solver to guide tableau search.
Rawson and Reger describe a literal-selection refinement that prefers extension choices whose selected literal is true in the current SAT assignment~\cite[\S~9.1]{satcop}.

\section{Research Vehicle}

\subsection{\cop{SATResetCoP} Specification}
\label{satresetcop}

We propose a new tableau prover based on \cop{SATCoP}, but guided by the goal of producing the ground clauses needed for a SAT refutation rather than using the \cop{leanCoP} heuristics that attempt to close the current tableau.

The resulting strategy is simple: aggressively reset the tree when the SAT solver stops receiving new ground clauses.
This adds a broader set of ground clauses to the solver in less time, allowing the SAT solver to reach unsatisfiability earlier.

\cop{SATResetCoP} uses many of the properties of \cop{SATCoP} described in the original paper and in its ``Future Research'' section~\cite[\S~9.1]{satcop}, but changes the tableau guidance to prioritise aggressive resets.

Like \cop{SATCoP}, the implementation keeps a persistent SAT solver alongside the active tableau search.
Ground clauses accumulated in earlier tableau attempts remain available in later attempts, and new clauses are added incrementally to the same solver instance.
This keeps the approach sound but incomplete.
However, we expect it to certify unsatisfiability of problems faster and in fewer steps.

\cop{SATResetCoP} randomises the \step{Start} and \step{Extension} steps in the same way as \cop{SATCoP}.
When starting a new tree, the implementation shuffles the possible starts and the literals inside each start clause.
When several extensions are possible from the leftmost open node, it shuffles the extension choices and the literal order inside the copied extension clause.

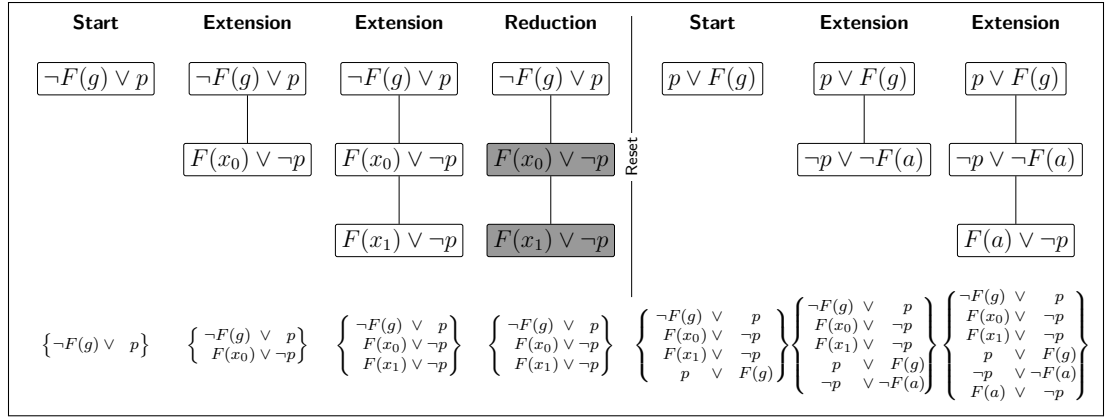
\begin{figure}[h]
	\centering
	\framebox[\linewidth][c]{\resizebox{\linewidth}{!}{\newcommand{\satdistance}{-160pt}
\newcommand{\dist}{75pt}
\newcommand{\bet}{5pt}

\renewcommand{\arraycolsep}{1pt}

\begin{tikzpicture}[
		attempt/.style = {
			every node/.style = {
				draw,
				rounded corners=1pt,
				inner xsep=3pt,
				inner ysep=2pt,
				font=\large
			},
			level distance=40pt,
			sibling distance=68pt,
			level 2/.style={sibling distance=63pt},
			edge from parent/.style={draw, -}
		},
		start/.style={},
		goal/.style={fill=black!40},
		bt/.style={edge from parent/.style={draw, dash pattern=on 2pt off 1pt on .5pt off 1pt}},
		title/.style={draw=none, font=\normalsize\bfseries},
		label/.style={draw=none, font=\footnotesize}
		]

\begin{scope}[attempt]
	\node[title] {\step{Start}};
	\node[start] at (0pt, -28pt) {$\neg F(g) \vee  p$};

	\node[label, align=center] at (0pt, \satdistance) {$
	\begin{Bmatrix}
		\neg F(g) \vee \phantom{\neg} p
	\end{Bmatrix}
	$};
\end{scope}

\begin{scope}[xshift={1*\dist}, attempt]
	\node[title] {\step{Extension}};
	\node[start] at (0pt, -28pt) {$\neg F(g) \vee  p$}
	child {
		node {$ F(x_0) \vee \neg p$}
	};

	\node[label, align=center] at (0pt, \satdistance) {$
	\begin{Bmatrix}
		\neg F(g) &\vee &\phantom{\neg} p \\
		\phantom{\neg} F(x_0) &\vee &\neg p
	\end{Bmatrix}
	$};
\end{scope}

\begin{scope}[xshift={2*\dist}, attempt]
	\node[title] {\step{Extension}};
	\node[start] at (0pt, -28pt) {$\neg F(g) \vee  p$}
	child {
		node {$ F(x_0) \vee \neg p$}
		child {
			node {$ F(x_1) \vee \neg p$}
		}
	};

	\node[label, align=center] at (0pt, \satdistance) {$
	\begin{Bmatrix}
		\neg F(g) &\vee &\phantom{\neg} p \\
		\phantom{\neg} F(x_0) &\vee &\neg p \\
		\phantom{\neg} F(x_1) &\vee &\neg p
	\end{Bmatrix}
	$};
\end{scope}

\begin{scope}[xshift={3*\dist}, attempt]
	\node[title] {\step{Reduction}};
	\node[start] at (0pt, -28pt) {$\neg F(g) \vee  p$}
	child {
		node[goal] {$ F(x_0) \vee \neg p$}
		child {
			node[goal] {$ F(x_1) \vee \neg p$}
		}
	};

	\node[label, align=center] at (0pt, \satdistance) {$
	\begin{Bmatrix}
		\neg F(g) &\vee &\phantom{\neg} p \\
		\phantom{\neg} F(x_0) &\vee &\neg p \\
		\phantom{\neg} F(x_1) &\vee &\neg p
	\end{Bmatrix}
	$};
\end{scope}

\draw (3.5*\dist+0.5*\bet, 3pt) -- (3.5*\dist+0.5*\bet, .85*\satdistance)
node[midway, fill = white, inner sep=2pt, rotate = 90] {\footnotesize \step{Reset}};

\begin{scope}[xshift={4*\dist+\bet}, attempt]
	\node[title] {\step{Start}};
	\node[start] at (0pt, -28pt) {$ p \vee  F(g)$};

	\node[label, align=center] at (0pt, \satdistance) {$
	\begin{Bmatrix}
		\neg F(g) &\vee &\phantom{\neg} p \\
		\phantom{\neg} F(x_0) &\vee &\neg p \\
		\phantom{\neg} F(x_1) &\vee &\neg p \\
		\phantom{\neg} p &\vee &\phantom{\neg} F(g)
	\end{Bmatrix}
	$};
\end{scope}

\begin{scope}[xshift={5*\dist+\bet}, attempt]
	\node[title] {\step{Extension}};
	\node[start] at (0pt, -28pt) {$ p \vee  F(g)$}
	child {
		node {$\neg p \vee \neg F(a)$}
	};

	\node[label, align=center] at (0pt, \satdistance) {$
	\begin{Bmatrix}
		\hspace{-2pt}\neg F(g) &\vee &\phantom{\neg} p \hspace{-2pt} \\
		\hspace{-2pt}\phantom{\neg} F(x_0) &\vee &\neg p \hspace{-2pt} \\
		\hspace{-2pt}\phantom{\neg} F(x_1) &\vee &\neg p \hspace{-2pt} \\
		\hspace{-2pt}\phantom{\neg} p &\vee &\phantom{\neg} F(g) \hspace{-2pt} \\
		\hspace{-2pt}\neg p &\vee &\neg F(a) \hspace{-2pt}
	\end{Bmatrix}
	$};
\end{scope}

\begin{scope}[xshift={6*\dist+\bet}, attempt]
	\node[title] {\step{Extension}};
	\node[start] at (0pt, -28pt) {$ p \vee  F(g)$}
	child {
		node {$\neg p \vee \neg F(a)$}
		child {
			node {$ F(a) \vee \neg p$}
		}
	};

	\node[label, align=center] at (0pt, \satdistance) {$
	\begin{Bmatrix}
		\hspace{-2pt}\neg F(g) &\vee &\phantom{\neg} p \hspace{-2pt} \\
		\hspace{-2pt}\phantom{\neg} F(x_0) &\vee &\neg p \hspace{-2pt} \\
		\hspace{-2pt}\phantom{\neg} F(x_1) &\vee &\neg p \hspace{-2pt} \\
		\hspace{-2pt}\phantom{\neg} p &\vee &\phantom{\neg} F(g) \hspace{-2pt} \\
		\hspace{-2pt}\neg p &\vee &\neg F(a) \hspace{-2pt} \\
		\hspace{-2pt}\phantom{\neg} F(a) &\vee &\neg p \hspace{-2pt}
	\end{Bmatrix}
	$};
\end{scope}

\end{tikzpicture}}}
	\caption{Example of \cop{SATResetCoP} on \step{SYN052-1.p}. \cop{SATCoP} and previous provers would backtrack after closing the branch with $F(x_1) \vee \neg p$; instead, \cop{SATResetCoP} resets into a new tree. This accumulates the ground clauses that form the UNSAT core shown at the bottom faster than in \cop{SATCoP}, allowing the SAT solver to reach unsatisfiability earlier. Substitutions are omitted for simplicity. At the end of the last extension, the accumulated ground clauses are unsatisfiable and the entire tableau prover can report UNSAT.}
	\label{reset_example}
\end{figure}

A new virtual operation, \step{Reset}, clears the tableau, the current substitution, and the proof sequence, and starts a new tableau attempt with a specified depth bound.
It does not clear the persistent SAT solver: every reset starts from an empty tree, while the SAT solver keeps the ground information produced by earlier attempts.
An example of how this operation works can be seen in \cref{reset_example}.

The grounded clauses used by \cop{SATResetCoP} come from clauses that have appeared in the tableau, after applying the tableau substitution accumulated so far.
Start clauses are added to the SAT state when the run is initialised.
After successful \step{Extension} and \step{Reduction} steps, the implementation grounds the clauses currently present in the tableau and records the genuinely new ones, which are later added to the SAT solver.

The grounding scheme follows the soundness argument of \cref{satcop}: the SAT solver receives only ground instances of clauses generated by the tableau.
Ground terms are kept as they are, and variables assigned by the current tableau substitution are replaced by their assigned terms.
Only tableau variables that remain unassigned need to be replaced by a ground term.
Our Python \cop{SATCoP}-style variants use the simpler fallback-style choice throughout, mapping every remaining tableau variable to the same fresh artificial constant.

The reset policy does not change this soundness argument.
Cuts and resets only affect which ground instances are discovered, and hence the completeness and performance of the search, not the soundness of a SAT refutation once found.

The aggressive reset-on-backtrack policy is the main implementation difference between \cop{SATResetCoP} and \cop{SATCoP}.
At a dead end, the prover first asks whether the current tableau attempt produced any new ground clauses that have not yet been added to the SAT solver.
If it did, those clauses are flushed to the SAT solver and the prover resets at the same depth.
This gives the next tableau attempt the same depth bound, but a larger SAT state and a fresh set of grounded clauses.
If the attempt did not produce any new ground clauses, the prover treats the current depth as exhausted and resets at the next depth instead of spending more steps on local backtracking.
The process stops as soon as the accumulated ground clauses become propositionally unsatisfiable.

Since \cop{SATResetCoP} resets the tree aggressively, it only increases the depth bound when the accumulated ground clauses have not been updated since the last \step{Reset}.
This prevents trees from growing out of control, while still trying to maximise the set of accumulated ground clauses.

Pseudocode for this implementation can be found in \cref{satresetcop_code}.

\begin{lstlisting}[float,floatplacement=h,caption={Pseudocode of the \cop{SATResetCoP} control policy.},label={satresetcop_code}]
def solve_tableau(start_clauses):
	solver := incremental_sat_solver()
	solver.add(ground(start_clauses))
	pending := empty_set

	max_depth := 1
	tableau := empty_tableau(max_depth)

	while solve_sat(solver) != UNSAT:
		model := current_model(solver)
		actions := legal_actions(tableau)

		if any Start, Extension, Reduction in actions:
			model_rank(a) :=
				2 if a is Start
				0 if selected_literal(a) is true in model
				1 if selected_literal(a) is unknown in model
				3 if selected_literal(a) is false in model

			next_action = argmin(actions, key = model_rank)
			tableau_step(next_action)

			if next_action is Reduction or Extension:
				pending.add(ground(tableau))

		else:
			if pending is not empty:
				solver.add_all(pending)
				pending := empty_set
				tableau := empty_tableau(max_depth)
			else:
				max_depth := max_depth + 1
				tableau := empty_tableau(max_depth)
\end{lstlisting}


\subsection{Implementation and benchmarking}

We implement \cop{SATResetCoP} using the \cop{Connections} library~\cite{connections_2023}, which provides primitives for implementing connection-tableau provers, together with the incremental SAT solver \cop{CaDiCaL}~\cite{cadical20}.
We also implement the control strategies \leancopc{}, \satcopzc{}, and \satcopc{} using the same backend, as mirrors of \cop{LeanCoP}, \cop{SATCoP\textsubscript{0}}, and \cop{SATCoP}.
Using a common backend makes the strategies easier to compare by keeping the tableau representation, rule applications, grounding interface, SAT interface, and step accounting fixed.

The source code for this implementation is available online,\footnote{\url{https://github.com/mfixman/resetsatcop}} and is released under the GNU Lesser General Public Licence~\cite{lgpl}.

We evaluate these 3 control strategies along with the test strategy \cop{SATResetCoP} on \num{7815} CNF problems and \num{8779} FOF problems from the Thousands of Problems for Theorem Provers (TPTP) library~\cite{tptp}.
The TPTP problems cover all 50 TPTP categories containing at least one CNF or FOF problem.

These benchmark sets provide a varied collection of problems for first-order provers.
TPTP also provides declared statuses for its problems, which are a useful sanity check for false unsatisfiability results.

We additionally test these strategies in the Bushy MPTP2078 library, which contains \num{2078} problems exported from the Mizar Mathematical library~\cite{mptp2078}.
This benchmark is useful because its problems share a large formal-mathematical vocabulary and test whether the strategies remain effective on related large-theory problems rather than only on the more heterogeneous TPTP collection.

\section{Experimental Evaluation}

\subsection{Experiment Design}

As a \cop{leanCoP}-style baseline we use \leancopc{}~\cite{leancopcut}.
This is a Python version of the \cop{leanCoP} Prolog implementation, with the two recommended predicates used in the first step of the schedule of \cop{LeanCoP~2.0}~\cite[\S~5.5]{leancopcut}.
\begin{description}
	\item[\texttt{cut}] Choose one extension or reduction and commit to it, similarly to Prolog's \texttt{!} operator.
	\item[\texttt{comp(7)}] The prover increases the depth bound by 1 whenever the current tree saturates, until reaching maximum height 7.
		Afterwards, the height limit is removed.
\end{description}

These simple policies tend to be very effective in tableau provers~\cite{leancopoverview}.

We add Python implementations in the connection prover of the two versions of \cop{SATCoP} presented by Rawson and Reger~\cite{satcop}: \satcopzc{}, corresponding to \cop{SATCoP\textsubscript{0}}, and \satcopc{}, corresponding to the version of \cop{SATCoP} with the improvements in the ``Future Research'' section that was submitted to CASC.
These implementations use similar techniques to the original provers, and since they use the same backend as \cop{SATResetCoP} allow for a fair comparison.

An overview of these strategies is shown in \cref{cops_table}.

\begin{table}[h]
	\centering
	\footnotesize
	\begin{tabularx}{\textwidth}{>{\sffamily}l X c c c c}
		\toprule
			Strategy & Explanation & Cut & SAT Refutation & SAT Guidance & Reset on Backtrack \\
		\midrule
			\leancopc{} & \cop{leanCoP} with initial set of predicates. & \checkmark{} & & & \\
		\addlinespace[1em]
			\satcopzc{} & Based on the initial implementation of \cop{SATCoP}~\cite{satcop}. & \checkmark{} & \checkmark{} & & \\
		\addlinespace[3pt]
			\satcopc{} & Improvements on \cop{SATCoP}~\cite[\S~9.1]{satcop}. & \checkmark{} & \checkmark{} & \checkmark{} & \\
		\addlinespace[1em]
			\cop{SATResetCoP} & Novel strategy, specified in \cref{satresetcop}. & \checkmark{} & \checkmark{} & \checkmark{} & \checkmark{} \\
		\bottomrule
	\end{tabularx}
	\caption{Control features of the baseline strategies and \cop{SATResetCoP}.}
	\label{cops_table}
\end{table}

Each problem-strategy pair is run until the problem is solved, or until either 120 seconds or 1 million steps pass, on a single core of the University of Cambridge CSD3 Cluster with dual-socket Intel Xeon Platinum 8368Q CPUs at 2.60 GHz, with 38 cores per socket and one thread per core, and 3370 MiB RAM per core~\cite{cambridge_csd3}.
We thank the Computer Science Lab of the University of Cambridge for providing this computer.

\subsection{Results}

\subsubsection{Individual Results}

\begin{figure}[h]
	\centering
	\begin{subfigure}[t]{.45\textwidth}
		\includegraphics[height=200pt]{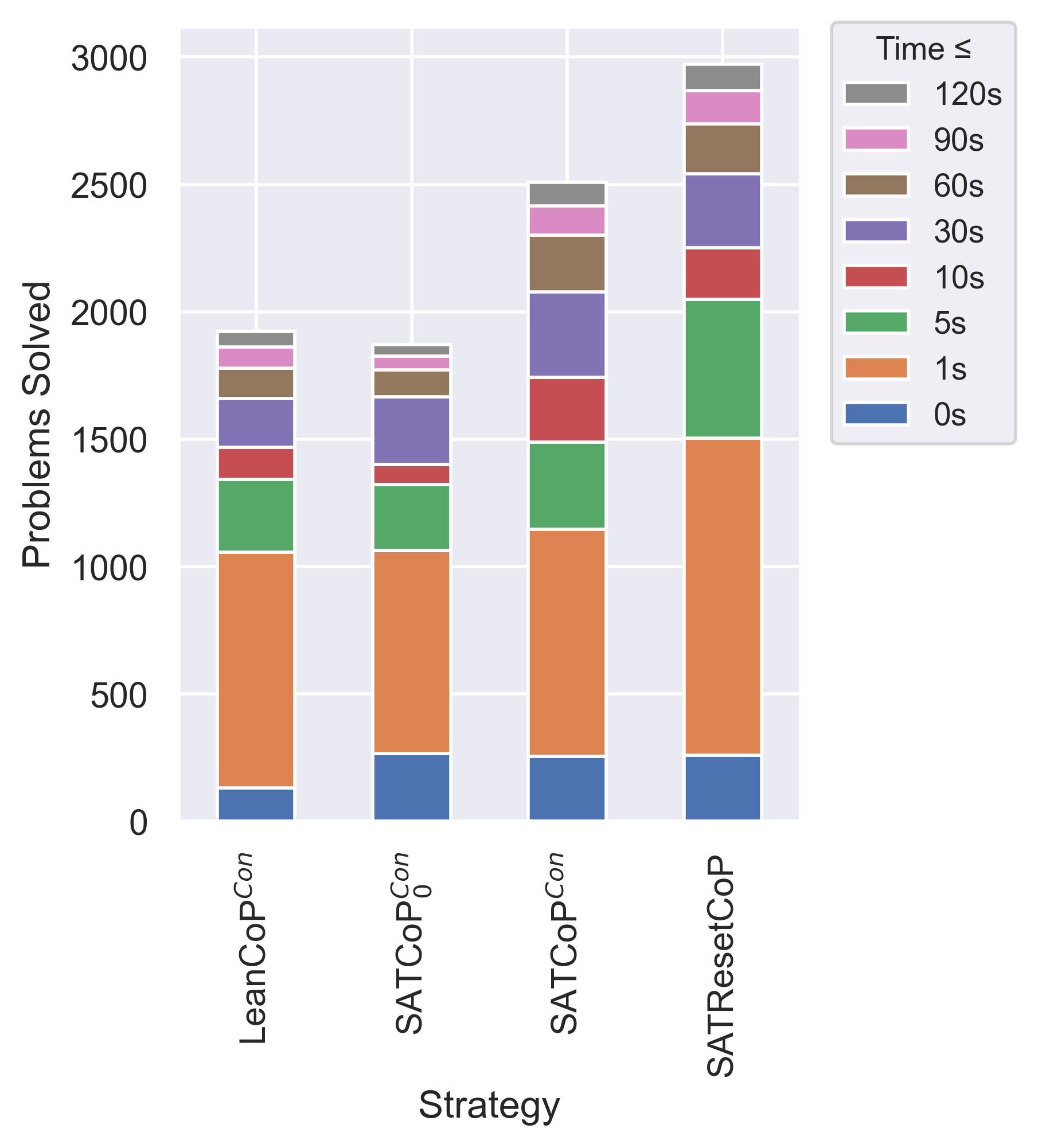}
		\caption{Problems solved, stacked by time in seconds.}
		\label{strategy_solved_time}
	\end{subfigure} \hspace{.05\textwidth}
	\begin{subfigure}[t]{.45\textwidth}
		\includegraphics[height=200pt]{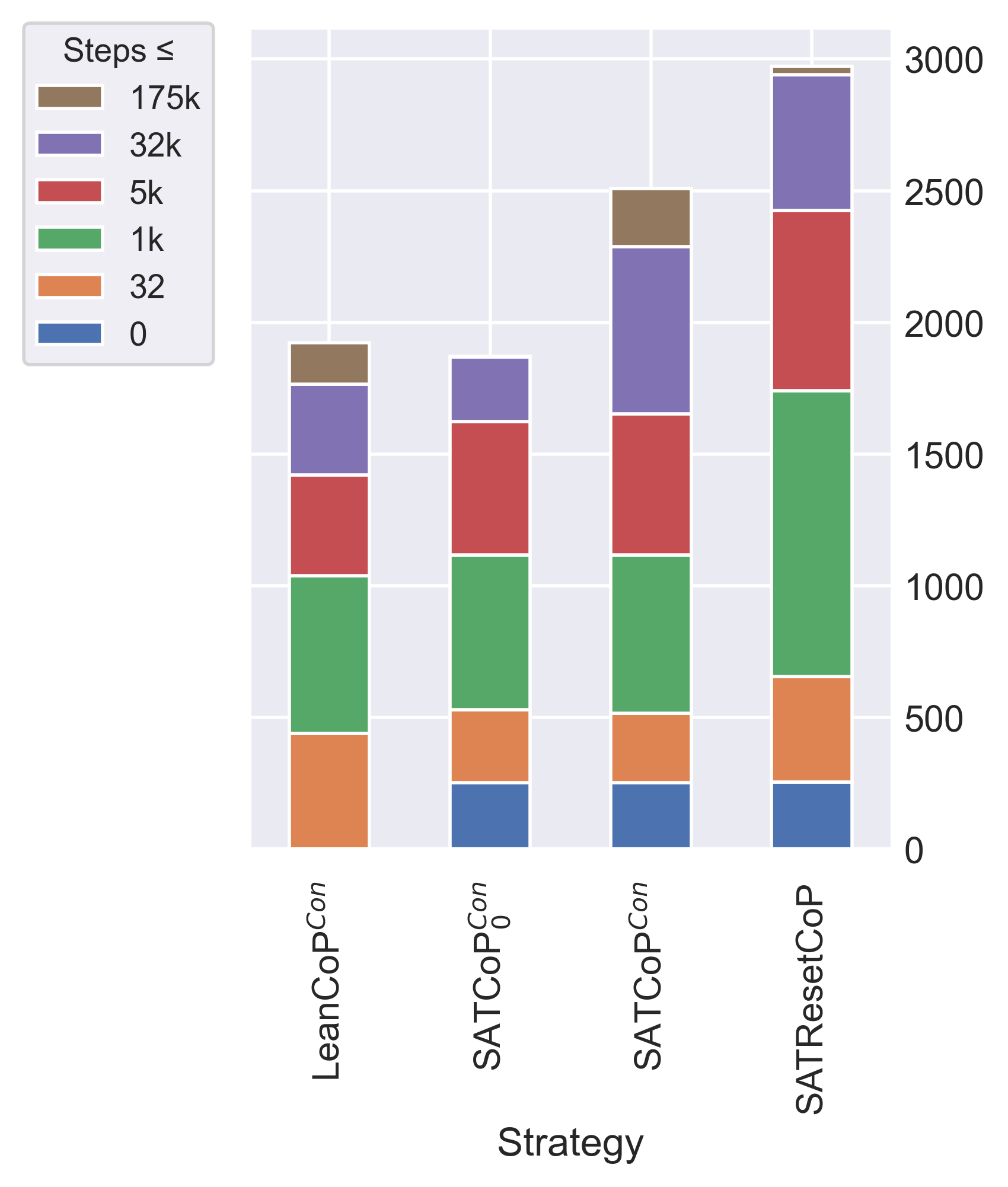}
		\caption{Problems solved, stacked by number of steps. SAT-based strategies can solve a problem in 0 steps if the initial ground SAT clause set is already unsatisfiable.}
		\label{strategy_solved_steps}
	\end{subfigure}
	\caption{Results of running the strategies on all TPTP problems under a time limit of 120 seconds and step limit of 1 million steps. \cop{SATResetCoP} solves more problems in fewer steps than in previous strategies, and also faster except for the niche category of $\leq 1s$ solves.}
\end{figure}

\cop{SATResetCoP} considerably improves on the number of problems solved by both \satcopzc{} and \satcopc{} under the 120-second time limit, as shown in \cref{problems_solved}.
The results are consistent with the claim that the SAT solver reaches unsatisfiability in less time and fewer \step{Start}, \step{Extension}, and \step{Reduction} steps than in the previous strategies, as seen in \cref{strategy_solved_time,strategy_solved_steps}.

The improvement is visible in both TPTP problem formats and on the MPTP2078-derived problems.
\cop{SATResetCoP} solves \num{529} more problems than \satcopc{}, with a net gain of \num{206} TPTP CNF problems, \num{256} TPTP FOF problems, and \num{67} Mizar Bushy problems.
Compared with \satcopzc{}, it solves \num{1395} more problems, an increase of \num{63.6}\%, and compared with \leancopc{} it solves \num{1251} more problems, an increase of \num{53.6}\%.

\begin{table}[h]
	\centering
	\footnotesize
	\begin{tabular}{>{\sffamily}l r r r r r r r r}
		\toprule
			& \multicolumn{2}{c}{TPTP CNF} & \multicolumn{2}{c}{TPTP FOF} & \multicolumn{2}{c}{MPTP2078} & \multicolumn{2}{c}{Total} \\
		\cmidrule(lr){2-3}
		\cmidrule(lr){4-5}
		\cmidrule(lr){6-7}
		\cmidrule(lr){8-9}
			strategy & Number & \% & Number & \% & Number & \% & Number & \% \\
		\midrule
			\leancopc{}       & 723  & 9.3  & 1200 & 13.7 & 413 & 19.9 & 2336 & 12.5 \\
			\satcopzc{}       & 804  & 10.3 & 1067 & 12.2 & 321 & 15.4 & 2192 & 11.7 \\
			\satcopc{}        & 1067 & 13.7 & 1442 & 16.4 & 549 & 26.4 & 3058 & 16.4 \\
			\cop{SATResetCoP} & 1273 & 16.3 & 1698 & 19.3 & 616 & 29.6 & 3587 & 19.2 \\
		\bottomrule
	\end{tabular}
	\caption{Number and percentage of TPTP CNF, TPTP FOF, and MPTP2078-derived problems solved within the time limit by each strategy. The total column combines all three benchmark sets.}
	\label{problems_solved}
\end{table}

\subsubsection{Comparative Results}
\label{comparative_results}

The tableau guidance in \cop{SATResetCoP} allows it to solve several problems left unsolved by strategies without the same mechanism for accumulating ground clauses, and even by \satcopzc{} and \satcopc{}, which use less aggressive guidance.
A comparison of newly solved and newly unsolved problems can be seen in \cref{newly_solved_table}.

This comparison is useful because the total number of solved problems hides changes in which problems are solved.
On TPTP, \cop{SATResetCoP} solves \num{598} problems that \satcopc{} does not, but loses \num{136} problems solved by \satcopc{}.
The reset policy is therefore changing the search distribution rather than just speeding up the same proof attempts.

\begin{table}[h]
	\centering
	\footnotesize
	\begin{tabular}{l r r r}
		\toprule
			Strategy & Newly Solved & Newly Unsolved & Total Solved \\
			\midrule
				\leancopc{}       & ---  & --- & 1923 \\
				\satcopzc{}       & 327  & 379 & 1871 \\
				\satcopc{}        & 661  & 23  & 2509 \\
				\cop{SATResetCoP} & 598  & 136 & 2971 \\
			\bottomrule
		\end{tabular}
	\caption{Problems from the TPTP dataset newly solved and newly unsolved by each strategy compared with the previous strategy in the table, together with the total number of solved problems. The reset policy changes which parts of the search space are explored: \cop{SATResetCoP} solves \num{598} TPTP problems missed by \satcopc{}, but misses \num{136} problems solved by \satcopc{}. This shows that aggressive resetting is not only a speedup of the same proof attempts, but a different search tradeoff.}
		\label{newly_solved_table}
\end{table}

The same tradeoff appears on the bushy \cop{MPTP2078} benchmark.
\cop{SATResetCoP} solves \num{111} MPTP2078 problems that \satcopc{} does not, while \satcopc{} solves \num{44} problems missed by \cop{SATResetCoP}.
The resulting net gain of \num{67} problems is smaller than on TPTP, but the pattern is similar.
Resets help when useful ground instances appear quickly, and hurt when the useful branch requires longer local tableau search.

This is the expected behaviour for a policy that treats the tableau mainly as a source of ground clauses.
When the early branches already expose the clauses needed by the SAT solver, resetting avoids spending many steps on similar local backtracking.
When the useful clauses only appear after a longer sequence of tableau choices, the same reset policy can discard the partial search state too early.

\begin{figure}[h]
	\centering\includegraphics[width=.45\textwidth]{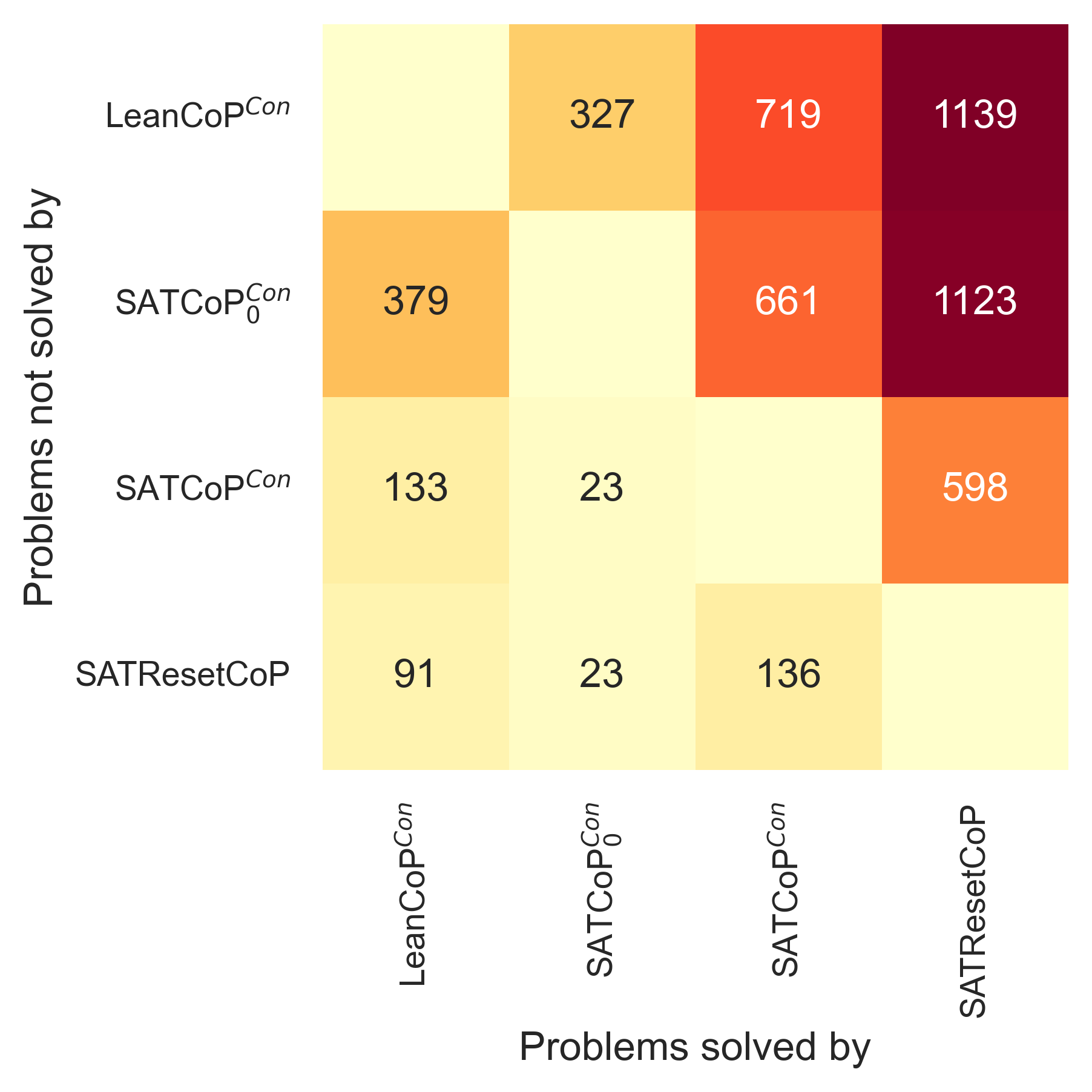}
	\caption{Pairwise comparison of strategy results: the cell in column \cop{y} and row \cop{x} shows the number of problems solved by strategy \cop{y}, but not by strategy \cop{x}. For example, \cop{SATResetCoP} solved 1123 TPTP problems that \satcopzc{} did not.}
	\label{solved_heatmap}
\end{figure}

\Cref{solved_heatmap} shows the number of problems solved by one strategy but not by another.

The improvement between \cop{SATResetCoP} and the best previous experimental strategy, our \satcopc{} implementation of \cop{SATCoP}, is especially visible in TPTP categories where the accumulated ground clauses become propositionally unsatisfiable quickly, such as \step{SYN}, \step{SWC}, and \step{SWV}.
This can be seen in \cref{problems_solved_categories}.

\begin{figure}[h]
	\centering
	\includegraphics[width=\textwidth]{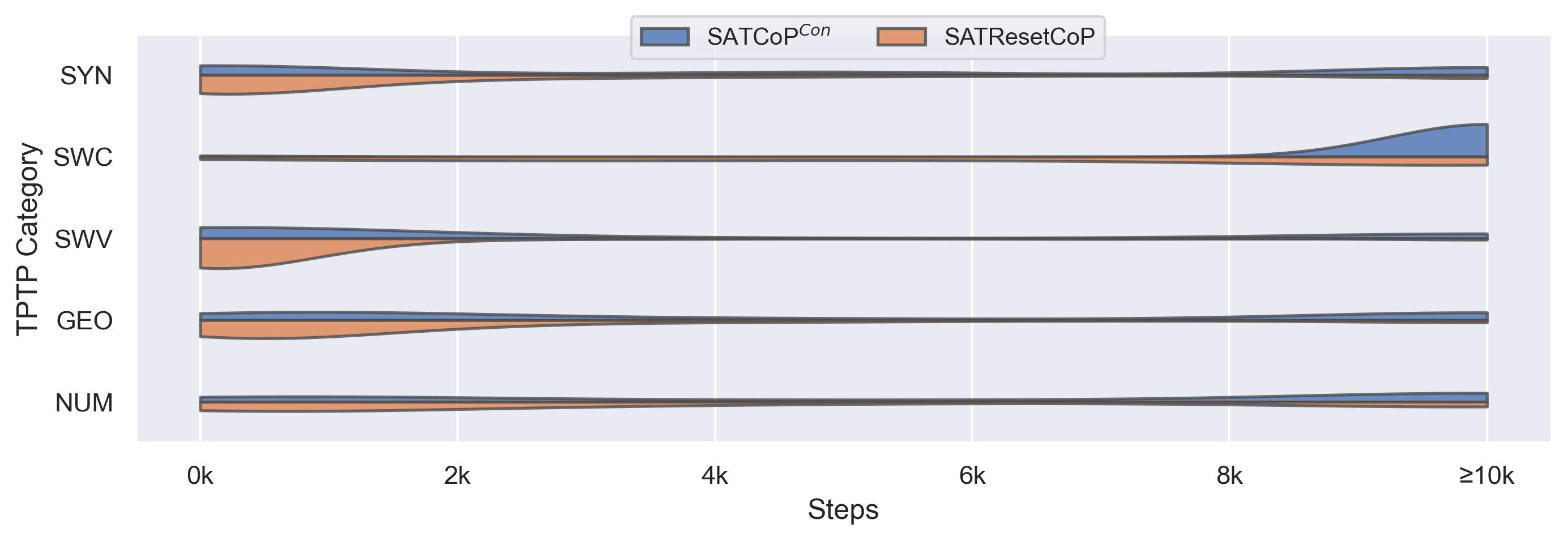}
	\caption{Density of problems solved at each step count by \satcopc{} and \cop{SATResetCoP} on the TPTP categories \cop{SYN}, \cop{SWC}, \cop{SWV}, \cop{GEO}, and \cop{NUM}. The area of each violinplot is the same; besides solving more problems in each category, \cop{SATResetCoP} solves many problems in fewer steps. Unsolved problems are collected in the $\geq 10k$ category on the right.}
	\label{violinplots}
	\vspace{10pt}%
	\includegraphics[width=\textwidth]{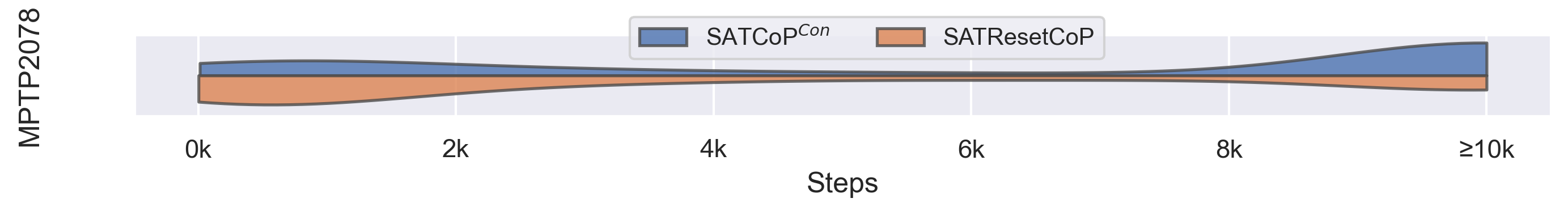}
	\caption{Density of problems solved at each step count by \satcopc{} and \cop{SATResetCoP} on the Mizar Bushy MPTP2078 dataset. Just like in TPTP, \cop{SATResetCoP} solves more problems in fewer steps. Unsolved problems are collected in the $\geq 10k$ category on the right.}
	\label{violinplot_mizar}
\end{figure}

As expected, this strategy also solves problems in fewer steps, as can be seen in \cref{violinplots} for TPTP problems and in \cref{violinplot_mizar} for the Mizar MPTP2078 dataset.

\begin{table}[h]
	\centering
	\footnotesize
	\begin{tabular}{l @{\hspace{1em}} r r r r r r @{\hspace{1em}} r}
		\toprule
			Strategy & SYN & SWC & SWV & GEO & NUM & Rest & Total \\
		\midrule
			\leancopc{}       & 486 & 14 & 163 & 216 & 86  & 958  & 1923 \\
			\satcopzc{}       & 459 & 2  & 161 & 177 & 90  & 982  & 1871 \\
			\satcopc{}        & 549 & 11 & 173 & 260 & 151 & 1365 & 2509 \\
			\cop{SATResetCoP} & 672 & 64 & 215 & 293 & 179 & 1548 & 2971 \\
		\bottomrule
	\end{tabular}
	\caption{Number of problems solved, per category, for each strategy.}
	\label{problems_solved_categories}
\end{table}

\subsubsection{Satisfiability Sanity Check}

The TPTP library includes declared statuses for its problems~\cite{tptp}.
While this does not demonstrate soundness, it provides a useful check that the strategy has not become unsound by mistake.

As a sanity check, we group the \cop{SATResetCoP} results by TPTP declared status in \cref{sat_proof}.
The results are consistent with what we would expect if the prover were sound: it does not find unsatisfiability (``Solved'') in any problem marked satisfiable or countersatisfiable.
It also does not solve any unknown or open problems, which would require solutions to be checked closely to ensure that they were not caused by an incorrect prover.

\begin{table}[h]
	\centering
	\footnotesize
	\begin{tabular}{lrrrr}
		\toprule
			& \multicolumn{2}{c}{FOF} & \multicolumn{2}{c}{CNF} \\
		\cmidrule(lr){2-3}
		\cmidrule(lr){4-5}
			Declared status & Unsolved & Solved & Unsolved & Solved \\
		\midrule
			Satisfiable         & 358  & 0    & 1095 & 0    \\
			CounterSatisfiable  & 746  & 0    & 0    & 0    \\
		\addlinespace[1ex]
			Unknown             & 139  & 0    & 745  & 0    \\
			Open                & 4    & 0    & 28   & 0    \\
		\addlinespace[1ex]
			Unsatisfiable       & 110  & 66   & 4674 & 1273 \\
			ContradictoryAxioms & 493  & 36   & 0    & 0    \\
			Theorem             & 5231 & 1596 & 0    & 0    \\
		\bottomrule
	\end{tabular}
	\caption{Number of problems solved and unsolved by \cop{SATResetCoP}, grouped by TPTP declared status. No satisfiable or countersatisfiable problem was incorrectly proven to be unsatisfiable.}
	\label{sat_proof}
\end{table}

\section{Conclusion}

The experiments on \cop{SATResetCoP} show that replacing local backtracking with resets can improve tableau performance when ground clauses are retained.

\Cref{comparative_results} shows that, on a 1 million step and 120 second time limit, \cop{SATResetCoP} solves \num{2971} problems.
This is \num{462} more problems than \satcopc{}, an 18\% increase, and \num{1048} more problems than \leancopc{}, a 54\% increase.
Pairwise, \cop{SATResetCoP} solves \num{598} problems that \satcopc{} does not, and \num{1139} problems that \leancopc{} does not.

The improvement is not uniform across TPTP.
It is strongest in categories where useful ground clauses accumulate quickly: \cop{SATResetCoP} solves \num{672} \step{SYN} problems, \num{64} \step{SWC} problems, and \num{215} \step{SWV} problems, compared with \num{549}, \num{11}, and \num{173} for \satcopc{}.
It also improves on \step{GEO} and \step{NUM}, solving \num{293} and \num{179} problems respectively.

The newly unsolved problems show the other side of this tradeoff.
These are cases where the accumulated ground clauses do not become propositionally unsatisfiable quickly enough, and the aggressive reset policy can discard tableau branches that a less aggressive prover eventually closes.

This suggests that accumulating ground clauses is a useful objective for connection tableaux, but not a replacement for all tableau control.
The most promising direction is therefore not a single fixed reset policy, but a schedule or controller that can use aggressive resets when ground instances are informative and fall back to ordinary connection search when they are not.

\section{Future Work}

\subsection{Schedule with different provers}
\label{schedule}

The reset policy is deliberately incomplete, and the newly unsolved problems show cases where less aggressive tableau search finds proofs that \cop{SATResetCoP} misses.
A natural next step is to use \cop{SATResetCoP} as one component of a schedule, following \cop{leanCoP 2.1} and the strategy-scheduling work of Mangla~\cite{leancopcut,mangla2022a}.
\cop{leanCoP 2.1}, for example, obtains stronger performance by scheduling several strategy variants with different options and resource limits, rather than relying on a single fixed configuration~\cite{leancopoverview}.

The simplest schedule would run \cop{SATResetCoP} first as a fast ground-clause generator, then switch to a more conventional \cop{leanCoP}- or \cop{SATCoP}-style strategy after a fixed time or step budget.
More adaptive schedules could use the SAT state itself, for example whether new ground clauses are still being produced or recent resets have stopped changing the accumulated clause set.

\subsection{Further Experiments}

The current implementation uses \cop{CaDiCaL} as its incremental SAT solver.
Future experiments should test whether the results are stable under different SAT solvers, solver restart policies, and incremental interfaces.
This matters because \cop{SATResetCoP} relies on the solver not only for the final unsatisfiability result, but also for the model information used to guide tableau extensions.

It would also be useful to vary the reset policy itself.
The present policy resets aggressively when the current tableau stops producing new ground clauses, but the newly unsolved problems suggest that some domains need longer tableau branches before the useful ground instances appear.
A broader experiment could compare fixed reset budgets, category-specific schedules, and policies that delay resets when recent SAT models still give useful literal guidance.

\subsection{CASC Submission}

\cop{SATResetCoP} will also be packaged as a CASC entrant.
This will test whether the improvements seen here survive the standard competition setting, including the official hardware, time limits, problem selection, and input requirements.
It will also make the comparison with earlier connection-tableau systems more direct, since both \cop{leanCoP} and \cop{SATCoP} have been evaluated through CASC.

Preparing a CASC version requires more than submitting the current research prototype.
The prover should expose a stable command-line interface, handle TPTP input robustly, report SZS statuses consistently, and include a reproducible installation path for the SAT backend.
This would turn the current experimental result into a system that can be compared against the broader ATP ecosystem.

\section{Acknowledgements}

This work was performed using resources provided by the Cambridge Service for Data Driven Discovery (CSD3) operated by the University of Cambridge Research Computing Service (\url{www.csd3.cam.ac.uk}), provided by Dell EMC and Intel using Tier-2 funding from the Engineering and Physical Sciences Research Council (capital grant EP/T022159/1), and DiRAC funding from the Science and Technology Facilities Council (\url{www.dirac.ac.uk}).

\bibliography{bibliography}

\end{document}